\documentclass{emulateapj} \usepackage{apjfonts,psfig} 

\shorttitle{VLT FLAMES spectra of ICPNe} 
\shortauthors{Arnaboldi, M. et al.} 
 
\begin{document} 
 
 
\title{The line-of-sight velocity distributions of intracluster 
planetary nebulae in the Virgo cluster core\footnote{Based on data 
collected with the FLAMES spectrograph at the UT2 of the VLT at Cerro 
Paranal, Chile, operated by ESO, during observing run 71.B-0147(A)}} 
 
 
\author{Magda Arnaboldi\altaffilmark{1},  
        Ortwin Gerhard\altaffilmark{2},  
        J. Alfonso L. Aguerri\altaffilmark{3},      
        Kenneth C. Freeman\altaffilmark{4}, \\
        Nicola R. Napolitano\altaffilmark{5}, 
        Sadanori Okamura\altaffilmark{6}, 
        Naoki Yasuda\altaffilmark{7}} 
\affil{$^1$ INAF, Oss. Astr. di Torino, Strada Osservatorio 20, 10025 
        Pino Torinese, Italy. arnaboldi@to.astro.it\\ 
       $^2$ Astronomical Institute, Univ. of Basel, Venusstrasse 7, CH-4102 
        Binningen, Switzerland. ortwin.gerhard@unibas.ch \\ 
       $^3$ Instituto de Astrofisica de Canarias, Via Lactea, 38200 La 
        Laguna, Spain. jalfonso@iac.es \\ 
       $^4$ RSAA, Mt. Stromlo Observatory, Cotter Road, Weston Creek, ACT 
        2611, Australia. kcf@mso.anu.edu.au \\ 
       $^5$ Kapteyn Astronomical Institute, P.O.B. 800, 9700 AV 
        Groningen, The Netherlands. nicola@astro.rug.nl\\ 
       $^6$ Dept. of Astronomy and RESCEU, School of Science, The
       Univ. of Tokyo, Tokyo 113-0033. okamura@astron.s.u-tokyo.ac.jp\\ 
       $^7$ Institute for Cosmic Ray Research, Univ. of Tokyo,
        Kashiwa, Chiba 277-8582. yasuda@icrr.u-tokyo.ac.jp  }
  
\begin{abstract} 
  Radial velocities of 40 intracluster planetary nebulae (ICPNe) in
  the Virgo cluster were obtained with the new multi-fiber FLAMES
  spectrograph on UT2 at VLT.  For the first time, the $\lambda$ 4959
  \AA\ line of the [OIII] doublet is seen in a large fraction (50\%)
  of ICPNe spectra, and a large fraction of the photometric candidates
  with m(5007) $\la 27.2$ is spectroscopically confirmed.
   
  ICPNe with the velocity dispersion of the Virgo cluster are found in
  our CORE field $1.^\circ2$ from M87. These may have originated from
  tidal mass loss of smaller galaxies in the M87 subcluster halo. In a
  field $0.25$ deg from M87, we see an extended stellar halo of M87 in
  approximate dynamical equilibrium, but with few ICPNe. Finally, in
  a field near M84/M86, the ICPNe velocities are highly correlated
  with the galaxy velocities, showing that any well-mixed intracluster
  population is yet to form. Overall, the measured velocity
  distributions confirm the non-uniform dynamical structure and
  on-going assembly of the Virgo cluster.
\end{abstract} 
 
 
\section{Introduction} 
After the serendipitous discovery of intracluster planetary nebulae
(ICPNe) in the Virgo (Arnaboldi et al. 1996) and Fornax (Theuns \&
Warren 1997) clusters, our group embarked on a systematic study of the
diffuse stellar population in the Virgo cluster.
The aim of this research is to measure the
fraction of cluster stars between galaxies, their radial
distribution, and their motions.  The discovery of ICPNe, the
subsequent narrow band imaging surveys to find ICPN samples (Feldmeier
et al.\ 1998, 2003a; Arnaboldi et al.\ 2002, 2003; Okamura et al.\
2002), and the HST observations of Virgo cluster empty fields
(Ferguson et al.\ 1998; Durrell et al.\ 2002) have revitalized the
study of diffuse intracluster light (ICL). The preliminary results on
the systematics of the ICL in different environments, from loose
groups to rich clusters (Gonzalez et al.\ 2000, Gal-Yam et al.\ 2003,
Castro-Rodriguez et al.\ 2003, Feldmeier et al.\ 2003b), and the
predictions from recent cosmological N-body and hydrodynamical
simulations (Napolitano et al.\ 2003,
Murante et al.\ 2004, Sommer-Larsen et al.\ 2004, Willman et al.\
2004) have proven the study of the ICL as a valuable tool to
investigate galaxy and galaxy cluster evolution.
 
ICPNe are the only component of the ICL whose kinematics can be
measured at this time. This is important since the high-resolution
N-body and hydrodynamical simulations predict that the ICL is
unrelaxed, showing significant substructure in its spatial and
velocity distributions in clusters similar to Virgo. While spatial
structures have been observed in the ICPN number density distribution
both in a single field (Okamura et al.\ 2002) and as field-to-field
variations (Aguerri et al.\ 2004), substructure in velocity space
still needs to be investigated.  

Early attempts by Freeman et al.\ (2000) with the AAT and 2dF
spectrograph provided only a few spectroscopically confirmed emission
line objects in the Virgo cluster core. Their identification as ICPNe
relies on the presence of the weaker [OIII] line ($\lambda$ 4959\AA)
in the summed spectrum of all the detected single-line
candidates. Because modern emission line candidate samples in Virgo
contain a modest fraction of Ly$\alpha$ emitters at $z=3.1$ (Aguerri
et al.\ 2004), spectral identification of ICPN either needs high
spectral resolution to resolve the typical broad, asymmetric
Ly$\alpha$ lines, and/or detection of the [OIII] $\lambda$ 4959/5007
\AA\ doublet\footnote{ The emission line sample of Kudritzki et al.\
(2000) was even dominated by Ly$\alpha$ emitters; however, the
luminosity function (LF) of this sample follows the LF of Ly$\alpha$
emitters in blank field surveys, not that of ICPNe in Virgo
(Castro-Rodriguez et al.\ 2003).}.
 
ICPN samples are sparse, with only a few tens of ICPNe in a 0.25
deg$^2$ field, and their fluxes are faint, from $1\times 10^{-16}$ to
$5\times 10^{-17}$ erg cm$^{-2}$ s$^{-1}$ in the [OIII] 5007\AA\ line.
Spectroscopic observations thus require 8 meter class telescopes and
spectrographs with a large field-of-view (FOV), of a fraction of a
square degree. The [OIII] emission lines from PNe are only a few km
s$^{-1}$ wide, so relatively high resolution spectra (R = 7000 to
10,000) are desirable to reduce the sky contamination.  The
FLAMES-GIRAFFE spectrograph on VLT, with its medium-high spectral
resolution, its FOV of 25 arcmin in diameter and 130 fibers in the
MEDUSA mode, is therefore very well suited to this project.
Here we present the results of the spectroscopic follow-up with FLAMES
of the ICPN candidates selected from three survey fields in the Virgo
cluster core. 
 
\section{Observations} 
Table~1 gives the three selected fields in the Virgo core, referred to
as FCJ, CORE, and SUB (Aguerri et al. 2004).  Spectra were acquired in
service mode (10 hrs were allocated to this observing run, 71.B-0147,
in priority A), with the FLAMES spectrograph at UT2 on VLT in the
GIRAFFE+MEDUSA configuration.  We used the low resolution grism LR
479.7, covering a wavelength range of 500~\AA, centred on 4797~\AA,
and a spectral resolution of 7500. This gives a velocity resolution of
40 km s$^{-1}$, and a typical velocity error of 12~km~s$^{-1}$. The redshifted
[OIII] emissions of ICPNe in the Virgo cluster core fall near the red
edge of the grism response.
 
The FLAMES FOV covers the FCJ field entirely, and a significant
fraction of both CORE and SUB. The total observing time was 2.5 hrs
for FCJ, and 2.6 hrs each for CORE and SUB; the exposure times
were based on the S/N estimate for detecting the [OIII] 5007\AA\ line
flux of $4.2 \times 10^{-17}$ erg cm$^{-2}$ s$^{-1}$, i.e. m(5007) =
27.2, with a S/N$\simeq 5$. Given that the [OIII] 4959/5007 \AA\
emission lines have relative intensities 1:3, we expect to be able
to detect the [OIII] 4959 \AA\ emission for the brighter candidates
only. The data reduction was carried out with the
GIRAFFE pipeline, for CCD pre-reduction, fiber identification,
wavelength calibration, geometric distortion corrections, co-addition
and extraction of the final 1D-spectra.

Comparing with earlier measurements by Freeman et al. (2000), the {\sc
rms} velocity difference for four ICPNe in common is 24~km~s$^{-1}$, where
their velocity error was 40~km~s$^{-1}$.

\begin{table}
\begin{center}
\caption{Observed fields and spectroscopic confirmation rates\label{tbl1}}
\begin{tabular}{lccc}
\tableline\tableline
\phantom{aaaaaaa} & FCJ\tablenotemark{a} & CORE\tablenotemark{a} & SUB\tablenotemark{a} \\
\tableline
$\alpha$(J2000) & 12:30:47.6  & 12:27:47.4 & 12:25:33.8 \\
$\delta$(J2000) & +12:38:32.4 & +13:21:40.1 & +12:47:01.2\\
$m_{lim}(5007)$\tablenotemark{b}    & 27.0 & 27.2 & 28.1 \\

$N_{\rm Fib}[<27.2]$\tablenotemark{b} & 18 & 34 & 18 \\

$N_{\rm PN}$\tablenotemark{b} & 15 & 12 & 13 \\

$N_{\rm PN}(\lambda 4959)$\tablenotemark{b} & 10 & 5 & 4 \\

$N_{\rm Fib}(<m_{min})$\tablenotemark{b} & 13 & 34 & 18 \\

$N_{\rm PN}(<m_{min})$\tablenotemark{b} & 11 & 11 &  13 \\

Rate$(<m_{min})$\tablenotemark{b} & 84\% & 32\%\tablenotemark{c}&  72\%\tablenotemark{d}\\

Predicted Rate$(<m_{min})$\tablenotemark{b}  & 80\% & 17\% & 100\%\tablenotemark{d} \\
\tableline
\end{tabular}
\tablenotetext{a}{References for photometric catalogues: Arnaboldi et
  al.\ (2002, FCJ), Aguerri et al.\ (2004, CORE), Okamura et al.\
  (2002, SUB).}  \tablenotetext{b}{Symbols denote: limiting magnitude
  of photometric sample, $m_{lim}(5007)$; number of fibers allocated
  to objects with $m<27.2$ and not in bad CCD regions, $N_{\rm
  Fib}[<27.2]$; number of detected PNe, $N_{\rm PN}$; number of
  detected PNe with double lines, $N_{\rm PN}(\lambda 4959)$; number
  of fibers allocated to objects brighter than min[$27.2$,
  $m_{lim}(5007)$], $N_{\rm Fib}(<m_{min})$; number of detected such
  PNe, $N_{\rm PN}(<m_{min})$; spectroscopic detection rate to
  $m_{min}$, Rate$(<m_{min})=N_{\rm PN}(<m_{min})/N_{\rm
  Fib}(<m_{min})$; corresponding rate predicted from photometry using
  simulations, Predicted Rate$(<m_{min})=$ number of PNe in
  statistically decontaminated sample divided by total number of
  candidates in the field (Aguerri et al.\ 2004).}
  \tablenotetext{c}{Higher than the predicted rate because in the
  fiber allocation priority was given to the brighter objects in the
  catalogue.}  
\tablenotetext{d}{All candidates are true ICPNe, based
  on detections in both [OIII] and H$\alpha$ narrow band
  images. Detected rate lower because of lower S/N.}
\end{center}
\end{table}

\section{Spectroscopic results}
\begin{figure} 
\plotone{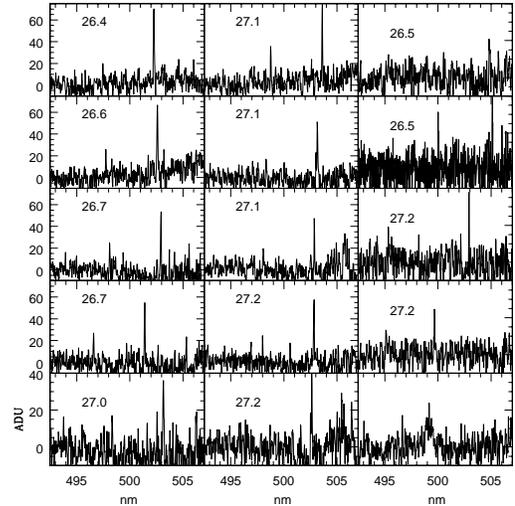} 
\caption{FLAMES spectra for 14 ICPNe observed in the FCJ (two left
columns) and CORE fields. The spectrum in the lower right corner is a
Ly$\alpha$ object, which shows a very broad line profile. The
$m(5007)$ magnitudes are marked on the individual frames.\label{fig1}}
\end{figure}

A total of 40 ICPN candidates were detected in the FLAMES spectra.  In
Fig.~\ref{fig1}, we show single ICPN spectra with the [OIII] 4959/5007
\AA\ doublet, and the resolved spectrum of a Ly$\alpha$ galaxy with a
broad asymmetric line.

In Table~1 we give results for the three pointings individually. In the
FCJ/CORE/SUB fields, we had 18/34/18 fibers allocated to sources
brighter or equal to 27.2 and in good regions of the CCD. In total, we
detected 15/12/13 sharp line emitters, and 0/2/0 Ly$\alpha$ emitters
which show one resolved asymmetric line.  The remaining spectra did
not show any spectral features in the wavelength range covered by
FLAMES. The fraction of confirmed spectra with both components of the
[OIII] doublet detected is 67\%/41\%/18\%.

The extracted 1D spectra for the CORE and SUB fields are noisier than
those for the FCJ, most probably because the selected reference stars
for the FLAMES MEDUSA configuration were not in the astrometric system
of the ICPN candidates, therefore fibers were not optimally
positioned.  For the SUB field, also the observing
conditions were slightly worse than for the other fields, resulting in
a low fraction of confirmed spectra with detected [OIII] doublet.
In the CORE field we checked that the summed spectrum of {\sl all}
sharp line emitters also shows the $\lambda$ 4959 \AA\ emission in
addition to the $\lambda$ 5007 \AA\ line. The line ratio is 3.5, but
with large error because of the noise in the summed spectrum: these
spectra are compatible with being all ICPNe.

Table~1 also gives the spectroscopic confirmation rates in the three
fields, for candidates down to the brighter of 27.2 or $m_{lim}(5007)$
from the photometry. This rate varies strongly from field to field
despite similar $m_{lim}(5007)$.  The most important reason for this
is different contamination of the samples by faint continuum stars,
erroneously classified as ICPNe because of a shallow off-band image.
Aguerri et al.\ (2004) investigated the contamination of the
photometric samples caused by faint continuum stars, [OII] emitters at
$z=0.347$, and high-z emitters, using simulations and blank field
surveys.  Table~1 shows the expected spectroscopic confirmation rates
based on these simulations. These are in close agreement with the
actual confirmation rates, showing that the photometric samples are
well understood. A high confirmation rate (small contamination) can be
achieved when the off-band image is sufficiently deep. See Aguerri et
al.\ (2004) for further details.

\section{ICPN Line-of-sight velocity distributions} 
 
\begin{figure} 
\plotone{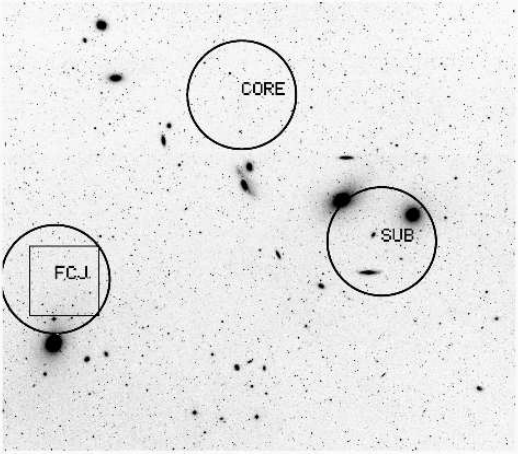} 
\caption{Pointings surveyed in the Virgo core region. The field shown
is $2^\circ \times 2^\circ$ with M87 near FCJ and M86/M84 near SUB. The
circles show the FLAMES FOVs. The square in the FCJ pointing marks the
area of the imaging surveyed field.\label{fig2}}
\end{figure}

\begin{figure} 
\plotone{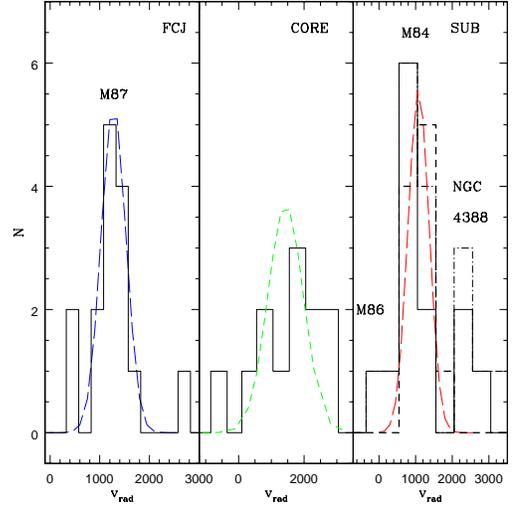} 
\caption{ICPN radial velocity distributions in the three pointings
(FCJ, CORE, and SUB). In the FCJ panel, the blue dashed line shows a
Gaussian with $\bar{v}_{rad} = 1276$ km s$^{-1}$ and $\sigma_{rad} =
247$ km s$^{-1}$. In the CORE panel, the green dashed line shows a
Gaussian with $\bar{v}_{rad} = 1436$ km s$^{-1}$ and $\sigma_{rad} =
538$ km s$^{-1}$, for VC galaxies dE and dS0 within 2$^\circ$ of M87
(from Binggeli et al. 1987). In the SUB sub, the overplotted dashed histogram
shows the radial velocities from a TNG spectroscopic follow-up (Arnaboldi et
al. 2003). The dashed red line shows a Gaussian with $\bar{v}_{rad} =
1079$ km s$^{-1}$ and $\sigma_{rad} = 286$ km s$^{-1}$ for the M84 peak. The
overplotted dashed-dotted lines show the SUB-FLAMES spectra including
those of HII regions, which have radial velocities in the M84 and
NGC~4388 velocity ranges.
\label{fig3}} 
\end{figure}

With these data, we can now for the first time determine radial
velocity distributions of ICPNe and use these to investigate the
dynamical state of the Virgo cluster. Fig.~\ref{fig2} shows an image
of the Virgo cluster core with the positions of our FLAMES
pointings. The heliocentric radial velocity histograms obtained from
the spectra in these fields are displayed in Fig.~\ref{fig3}.  Clearly
the histograms for the three pointings are very different.

In the FCJ field, the velocity distribution of the PNe is not
consistent with a single Gaussian with either the fitted velocity
dispersion $\sigma_{\rm FCJ}=573$ km s$^{-1}$, or the more canonical Virgo
$\sigma=800$ km s$^{-1}$, based on a $\chi^2$-test. Instead, it is
dominated by a narrow peak, with $\bar{v}_{p} = 1276\pm 71$ km
s$^{-1}$ and $\sigma_{p} = 247\pm 52$ km s$^{-1}$, which we identify
with the halo of M87 below. In addition, there are 3 outliers, 2 at
low velocity. All three are unusually bright, in the bright fall-off
of the M87 PN luminosity function (PNLF), and they shift the FCJ luminosity
function to a brighter cutoff (Arnaboldi et al.\ 2002).  The ICL
surface brightness associated with these 3 outliers, e.g.\ the likely
true ICPNe in the FCJ field, is $\mu_B \simeq 30.7$ mag arcsec$^{-2}$
if we use the conversion for M31 from Ciardullo et al.\ (1989),
$\alpha_{1,B}$.  This is comparable with the surface brightness
measurements of Ferguson et al.\ (1998) and Durrell et al.\ (2002)
from intracluster red giants in fields further away from M87.
 
The M87 peak of the FCJ velocity distribution contains 12 velocities,
which are described well by a Gaussian according to KS and $\chi^2$
tests. Their luminosity function is consistent with the PNLF in the
inner $4'$ of M87 (Ciardullo et al.\ 1998), with a KS probability
$p_{KS}=0.46$.  The average velocity agrees with that of M87,
$v_{sys} = 1307$ km s$^{-1}$. The value of $\sigma_p$ is consistent
with the stellar velocity dispersion profile extrapolated outwards
from $\simeq 150''$ in Figure~5 of Romanowsky \& Kochanek (2001) and
falls in the range spanned by their dynamical models for the M87 stars
(the center of FCJ is $15.'0\simeq\,65$ kpc from M87,
for an assumed M87 distance of $15$ Mpc).  To infer the surface
brightness corresponding to the 12 PNe in the M87 peak, we again use
$\alpha_{1,B}$ for M31, because of the colour gradient observed in the
outer parts of M87 (Goudfrooij et al.\ 1994). The resulting
$\mu_B=29.2$ mag arcsec$^{-2}$ falls on the average M87 surface
brightness profile extrapolated from $\simeq 400''$.  $\mu_B$ and
$\sigma_p$ are thus also consistent with the extrapolated dynamical
model of Romanowsky \& Kochanek (2001) for the M87 stars, in the
distribution of dark matter inferred by them, which is also similar to
that determined by Matsushita et al.\ (2002).  By contrast, the
available data and dynamical models show an approximately flat
dispersion profile for the globular clusters in M87, at $\sigma_{\rm
GC}\simeq 350$ km s$^{-1}$, corresponding to a shallower radial
distribution.  The main result from our measurement of $\sigma_p$ is
that M87 has a stellar halo in approximate dynamical equilibrium out
to at least $65$ kpc.
 
In the CORE field, the distribution of ICPN line-of-sight (LOS)
velocities is clearly broader than in the FCJ field and consistent
with a Gaussian ($p_{KS}=0.9$). It has $\bar{v}_{C} = 1491\pm 290$ km
s$^{-1}$ and $\sigma_{C} = 1000\pm 210$ km s$^{-1}$; the median is
$1791$ km s$^{-1}$.  The CORE field is in a region of Virgo devoid of
bright galaxies, but contains 7 dwarfs, and 3 low luminosity E/S near
its S/W borders. None of the confirmed ICPNe lies within a circle of
three times half the major axis diameter of any of these galaxies, and
there are no correlations of their velocities with the velocities of
the nearest galaxies where these are known. Thus in this field there
is a clear IC stellar component.
 
The mean velocity of the ICPN in this field is similar to that of
25 Virgo dE and dS0 within 2$^\circ$ of M87, $<v_{\rm dE,M87}> =
1436\pm108$ km s$^{-1}$ (Binggeli et al.\ 1987), and with that of 93
dE and dS0 Virgo members, $<v_{\rm dE,Virgo}> = 1139\pm67$ km s$^{-1}$
(Binggeli et al.\ 1993).  However, the velocity dispersion of these
galaxies is smaller, $\sigma_{\rm dE,M87}=538\pm 77$ km s$^{-1}$ and
$\sigma_{\rm dE,Virgo}=649\pm 48$ km s$^{-1}$.
 
The inferred luminosity from the ICPNe in the CORE field is $1.8\times
10^9 L_{B,\odot}$ (Aguerri et al.\ 2004). This is about three times
the luminosity of all dwarf galaxies in this field, $5.3\times 10^8
L_{B,\odot}$, but an order of magnitude less than the luminosities of
the three low-luminosity E/S galaxies near the field borders. Using
the results of Nulsen \& B\"ohringer (1995) and Matsushita et al.\
(2002), we estimate the mass of the M87 subcluster inside 310 kpc (the
projected distance $D$ of the CORE field from M87) as $4.2\times
10^{13} M_\odot$, and compute a tidal parameter $T$ for all these
galaxies as the ratio of the mean mass density within $D$ to the mean
density of the galaxy. We find $T=0.01-0.06$, independent of galaxy
luminosity.  Since $T\sim D^{-2}$, any of these galaxies whose orbit
{\sl now} comes closer to M87 than $\sim 60$ kpc would be subject to
severe tidal mass loss.  Thus the ICPN population in the CORE field
could be debris from the tidal disruption of small galaxies on nearby
orbits in the M87 halo. The relatively small number of ICPNe in the
FCJ field at $D=65$ kpc could then mean that most of the tidally
disrupted galaxies did not orbit as deep into M87.

In the SUB field the velocity distribution from FLAMES spectra is
again different from CORE and FCJ. The KS test gives low
probabilities, $p_{KS}=0.13$ and $p_{KS}=0.03$, that the SUB histogram
could be drawn from the velocity histogram in the CORE field or the
Gaussian fitted to this, respectively. Instead, the SUB histogram of
LOS velocities shows substructures that are highly correlated with the
systemic velocities of M86, M84 and NGC 4388.  The association with
the three galaxies is strengthened when we plot the LOS velocities of
4 HII regions (see Gerhard et al.\ 2002) detected with FLAMES in this
pointing.  The highest peak in the distribution coincides with M84,
and even more so when we add the LOS velocities obtained previously at
the TNG (Arnaboldi et al.\ 2003).  The 10 TNG velocities (2 velocities
were measured in addition to the Arnaboldi et al. (2003) sample) give
$\bar{v}_{\rm M84} = 1079\pm 103$ km s$^{-1}$ and $\sigma_{\rm M84} =
325\pm75$ km s$^{-1}$ within a square of $4 R_e \times 4 R_e$ of the
M84 center.  The 8 FLAMES velocities in the M84 subpeak give
$\bar{v}_{\rm M84} = 891\pm 74$ km s$^{-1}$ and $\sigma_{\rm M84} =
208\pm54$ km s$^{-1}$, going out to larger radii.  Note that this
includes three over-luminous PNe not attributed to M84 previously. The
combined sample of 18 velocities gives $\bar{v}_{\rm M84} = 995\pm 69$
km s$^{-1}$ and $\sigma_{\rm M84} = 293\pm50$ km s$^{-1}$.  Most
likely, all these PNe belong to a very extended envelope around M84
(see the deep image in Arnaboldi et al.\ 1996). It is possible that
the somewhat low velocity with respect to M84 may be a sign of tidal
stripping by M86, or of a recent merger with a smaller galaxy. In the
latter case, the overluminous PNe might be due to a younger or a more
metal-poor population (Dopita et al.\ 1992).

\section{Conclusions}

We have presented the first measurements of the velocity distribution
of ICPNe in three fields of the Virgo cluster. Overall, these
velocity measurements confirm the view that Virgo is a highly
non-uniform and unrelaxed galaxy cluster, consisting of several
subunits that have not yet have had time to come to equilibrium in a
common gravitational potential.

A well-mixed IC stellar population is seen clearly only in the CORE
field, in the outer parts of the M87 subcluster. Here the velocity
distribution is consistent with a single cluster Gaussian, and the
ICPNe might well have their origin in the tidal effects of the halo of
this subcluster on its galaxy population.  In the SUB field near M84
and M86, the ICPNe do not appear virialized; their velocities are highly
correlated with those of the large galaxies in the field.  In fact,
there are regions in Virgo where no ICPNe are found (the LPC field of
Aguerri et al.\ 2004).

The measurements have also shown that M87 has a very extended envelope
in approximate dynamical equilibrium, reaching out to at least $65$
kpc. The cluster ICPN population in our FCJ field has density comparable 
with that in other fields further out, indicating a shallow ICL density
distribution.


\acknowledgments M.A. would like to thank L. Pasquini and F. Primas
for their help during the VLT-FLAMES observing block preparation and
service observing, and C. Cacciari, E. Pancino and E. Rossetti for
their help with the GIRAFFE pipeline. We
thank B. Binggeli and M. Capaccioli for useful discussions,
B. Binggeli for providing his Virgo cluster galaxy catalogue with the
dE \& dS0 radial velocity data, and an anonymous referee for
constructive comments. We acknowledge financial support by INAF $-$
Projects of National Interests (P.I. MA), SNF grant 200020-101766,
and Spanish DGES grant AYA2001-3939. NRN is supported by
a MC fellowship within the EC Fifth Framework Program.  This
research has made use of the NASA/IPAC Extragalactic Database (NED).


\end{document}